\documentclass[12pt]{iopart}
\usepackage{iopams}
\usepackage{graphicx,latexsym,amssymb,cite} 
\begin{document}

\title[Critical Random Matrix Ensemble]{Energy Level Statistics of a Critical Random Matrix Ensemble}

\author{Macleans L. Ndawana\dag\
\footnote[3]{Permanent address: University of Zambia, Department of Physics, Lusaka, Zambia}
and Vladimir E. Kravtsov\ddag  
\footnote[4]{Landau Institute for Theoretical Physics, Kosygina str. 2, 117940 Moscow, Russia}
}

\address{\dag\ Institut f\"{u}r Physik, Technische Universit\"{a}t Chemnitz, 
D-09107 Chemnitz, Germany} 

\address{\ddag\ International Center for Theoretical Physics, P. O. Box 586, 34100 Trieste, 
Italy}

\begin{abstract}
We study energy level statistics of a critical random matrix ensemble of 
power-law banded complex Hermitian matrices. We compute the level compressibility via the 
level-number variance and compare it with the analytical formula for the exactly solvable model of
Moshe, Neuberger and Shapiro. 
\end{abstract}

\section{Introduction}
\label{int}
Random matrix ensembles (RME) provide a natural frame work for the statistical
description of quantized energy levels of complex systems. This is probably best 
explained by paraphrasing Dyson \cite{Dys62}: in ordinary statistical mechanics a 
renunciation of knowledge about the system is made {\em i.e.,} by assuming that all states 
of a large ensemble are equally probable. Therefore only the overall behavior of the 
system is accessible. Thus orthodox statistical mechanics is clearly inadequate for 
the discussion of energy level statistics  of complex systems. In energy level
statistics one wishes to make statements about the fine detail of the energy 
level structure, and such statements cannot be made in terms of an ensemble of 
states. Random matrix theory (RMT) turns out to be a suitable tool in 
making such statements about the eigen-spectrum of complex systems {\em e.g.,} energy 
levels of highly excited nuclei \cite{Wig51}. The underlying assumption in RMT is 
that one renounces exact knowledge {\em not} of the state but of the {\em nature} of 
the system itself. This is the fundamental difference between ordinary statistical 
mechanics and energy level statistics.

There are three RME that emerge from an analysis of real eigen-spectra of random Hermitian 
matrices \cite{Meh90}; the Gaussian orthogonal ensemble (GOE) for a system with 
time-reversal without spin-$\frac{1}{2}$ interactions, the Gaussian symplectic ensemble 
(GSE) for a system with time-reversal and spin-$\frac{1}{2}$ interaction, and the Gaussian 
unitary ensemble (GUE) for a system without time-reversal. These ensembles (generally called
Wigner-Dyson (WD) statistics) exhibit strong energy-level correlations leading 
to the phenomenon of level repulsion \cite{Haa92} in their respective eigen-spectra. One could
add the Poisson statistics (PS) as the fourth ensemble which does not exhibit level repulsion in 
it's corresponding eigen-spectra.

In general, the WD statistics describe a system of delocalized states and the PS describe 
a system of localized states. The transition between these two cases controlled by the strength of
disorder ({\em i.e} the Anderson model \cite{And58}) is a quantum critical phenomenon. At a critical 
disorder the states acquire a special property of {\it multifractality} \cite{KraM97,Jan98} which marks 
a qualitative difference as compared to truly delocalized states in a metal and truly localized states in 
an insulator. These critical states correspond to the {\em critical level statistics} (CLS) \cite{Shkl,KLAA}. 

There are numerous suggestions in literature
\cite{KraM97,MosNS94,MirFDQ96,MutCIN93,Mir00,Jan98,BogBP97,GarV00,BleCM94,GarV02} describing 
CLS by ensembles of random matrices. Initially, there were two main requirements for the critical RME: 
the first was that the level statistics be system-size independent and secondly, the ensemble 
should be able to interpolate between WD and PS \cite{Shkl}. Later on it was realized \cite{ChaKL96,KraM97} that 
the key property of the critical states relevant for the CLS is multifractality. The power-law banded RME 
suggested in \cite{MirFDQ96} emerged as the best candidate for the critical RME as its level statistics 
is known to possess these two properties and its eigenfunction statistics is multifractal. Thus it has 
been suggested \cite{KraM97} to be the critical random matrix ensemble (CrRME). 
In the limit of large bandwidth this model turns out to be the best deformation of the WD-RME that 
generates correlations in the eigenstates but retains all the basic properties of WD theory,
including the factorizability of higher spectral correlation functions and mapping onto the 
Calogero-Sutherland model of fictitious interacting fermions \cite{KTs}.

In the absence of an exact solution of CrRME various mappings on and approximate correspondences 
with exactly solvable models have been used in order to study the properties and CrRME. Here we 
mention the mapping onto the nonlinear sigma-model \cite{MirFDQ96}, and the approximate equivalence 
\cite{KraM97} between the CrRME and the exactly solvable RME suggested in Refs.\cite{MutCIN93,MosNS94}. 
Making use of the exact solution \cite{MutCIN93} it was possible \cite{Nish} to compute the level spacing
distribution function (LSDF) for this model and compare it with the exact diagonalization of the 3d Anderson 
model for the critical value of disorder. The coincidence of results (for certain values of the parameter $b$
which were found from the fitting of the far tails of LSDF) was amazing \cite{Nish,Can96}: in all three Dyson 
symmetry classes the deviation was not larger than between the Wigner surmise and the exact WD distribution 
function.

In this paper we study the level number variance in the CrRMT suggested \cite{MirFDQ96}.
The ultimate goal here is to compare the numerical simulations of this critical ensemble 
and the analytical results for the exactly solvable ensemble in Ref. \cite{MosNS94} (which will be
referred to as MNS) for all values of the control parameter $b$. For large $b\gg 1$ (weak multifractality) these 
two ensembles are known to be equivalent \cite{KraM97}. The first correction to the Poisson spectral  statistics 
at small $b\ll 1$ (strong multifractality) is also the same \cite{MirE00,EveM00}. We would like to check these
predictions numerically for the unitary CrRME and to see how large are the deviations between these two RME 
for $b\sim 1$.

The paper is organized as follows: in Section \ref{spec} we give a general overview of the 
level-number variance ($\Sigma_2$-statistics) and the calculation of the level 
compressibility, in Section \ref{resdis} we give some numerical details and finally a summary 
and outlook of this work is given in Section \ref{conc}.
%
%
\section{$\Sigma_2$-Statistics and Level Compressibility}
\label{spec}
Our model is the power-law random banded matrix defined as an ensemble of random
Hermitian $N \times N$ matrices belonging to GUE. The real and imaginary parts of 
the matrix elements are independently distributed Gaussian random variables with 
zero mean $\langle H_{ij}\rangle=0$ and the variance of the matrix elements is given by;  
\begin{equation}
\label{vari}
\langle \left|H_{ij}\right|^{2}\rangle= \ 
\left[1+\frac{1}{b^2}\frac{\sin^{2}(\pi \left|i-j\right|/N)}{(\pi/N)^2}\right]^{-1} \
\times
\left\{ \begin{array}{ll}
\frac{1}{\beta} & \textrm{if $i=j$} \\
\frac{1}{2} & \textrm{if $i \ne j$}
\end{array} \right.
\end{equation}
where $b$ is a parameter that controls the fractal dimensions of the critical eigenfunctions 
and $\beta$ is the symmetry parameter {\em i.e.,} $\beta=1,2$ and $4$ for the GOE, GUE 
and GSE respectively. This model remains critical for any arbitrary value of $b$.

From a physical point of view the GUE can be realized by putting a system in a strong 
external magnetic field {\em i.e.,} provided the splitting of levels by the magnetic 
field is of the same order of magnitude as the average level spacing in the absence of 
the field.  

In this paper we study the level compressibility $\chi(N,b)$ via the 
level-number variance $\Sigma_{2}\left(\left< n \right>\right)$ of the above model. The level-number 
variance is a statistical quantity that provides a quantitative measure of the long-range "rigidity" 
of the energy spectrum \cite{Meh90}; 
\begin{equation}
\label{vari2}
\Sigma_{2} \left(\left<n\right> \right)=\left<n^2\right>-\left<n\right>^2
\end{equation}
The behavior of $\Sigma_2$ is known (as a function of the mean number of levels 
$\left< n \right> \gg 1$) in the delocalized and localized phases respectively. However, in the 
critical regime, the level-number variance has been conjectured to be Poisson-like \cite{AltZKS88,ChaKL96};
\begin{equation}
\label{eq-sigd}
\Sigma_{2} \sim \left\{ \begin{array}{ll}
{\frac{2}{\pi^{2}\beta}}\ln \left(\left<n\right> \right) & \textrm{delocalized,} \\
\quad \chi \left <n \right> & \textrm{critical,} \\
\quad \left< n \right> & \textrm{localized,}
\end{array} \right.
\end{equation}
where the level compressibility $\chi$ is another important parameter to characterize the 
localization-delocalization transition (LDT) and takes values $0\le\chi\le1$, being zero in 
the delocalized state and unity in the localized state. Formally, one computes $\chi$ as;
\begin{equation}
\label{eq-si3}
\chi \approx \lim_{\left< n \right> \rightarrow \infty} \
\lim_{N\rightarrow \infty} \frac{d\Sigma_2\left(\left< n\right>\right)}{d\left< n \right>}
\end{equation}
The limits in Eqn.(\ref{eq-si3}) do not commute and their non-commutativity is attributed to the
violation of the normalization sum rule \cite{ChaKL96}. It is remarkable (see 
\cite{Kra96,KraM97,ChaKL96,ChaLS96} for details) that  $\chi$ can be expressed in terms of the
fractal or the correlation dimension $D_2$ and the spatial dimension $d$. In Ref.\cite{ChaKL96} 
an explicit and very simple relationship between the critical spectral and eigenfunction 
statistics has been suggested: 
\begin{equation}
\label{fract}
\chi=\frac{1}{2}\left(1-\frac{D_2}{d}\right)
\end{equation}
There is vast analytical and numerical evidence that this formula is correct at weak
multifractality. However, its status for the case of strong multifractality (small $D_{2}$) 
raises much doubts \cite{EveM00}. We later see that whilst our numerical data are in agreement 
with Eqn.(\ref{fract}) in the limit $b^{-1} \rightarrow 0$ they are in disagreement \cite{MirE00,EveM00}
for the opposite limit. At the same time the overall agreement with the analytical result Eqn.(10) for 
the MNS model is unexpectedly good for all values of $b$ and the moderately large matrix sizes 
up to $N=400$. We were not able to detect any systematic dependence on the size of the matrix 
$N$ in this range.

In the next section we turn to the numerical details of how to compute the level compressibility.
%
%
\section{Results and Discussion}
\label{resdis}
We performed numerical simulations of the RME described by Eqn.(\ref{vari}) for values of 
$b^{-1}\in\left[0.4,40\right]$. The matrices are not sparse and as such the use of fast and efficient 
algorithms {\em i.e.,} the Lanczos algorithm is not possible. However, we calculated the spectra 
of $N \times N$ matrices for $N=50$ with $10000$ realizations to  $N=400$ with $2000$ 
realizations by exact numerical diagonalization and hence accumulating $\sim 1\times 10^{5}$ eigenvalues
for each value of $N$. The results we discuss here are for the GUE only.

The average density of states as a function of energy $E$ is well described by;
\begin{equation}
\label{density}
\rho \left(E \right)=\frac{N}{{\pi^2}b}\int_{0}^{+\infty} \
\frac{dt}{1+\left(1/z\right) \exp \left[\left(\beta/2\right) \left(E^{2}+t^{2}\right)\right]}
\end{equation}
where $z=\exp\left(\pi \beta b\right)-1$. The two limiting cases of Eqn.(\ref{density}) are;
\begin{equation}
\label{semi}
\rho \left( E \right)= \left \{\begin{array}{ll}
\frac{N}{{\pi^2}b}\sqrt{2\pi b - E^2} & \textrm{Semi-circle if $\pi \beta b\gg 1$} \\
{N\sqrt{\frac{\beta}{2\pi}}}\exp \left(-\beta E^{2}/2 \right) & \textrm{Gaussian otherwise} \end{array}
\right.
\end{equation}
All our calculations are done at the band center (containing $\left< n\right> \gg 1$) where 
the density of states is nearly constant.   
The calculation of $\chi$ from $\Sigma_2$ is non-trivial as there are finite-size effects 
$\sim -\left< n\right>^{2}/N$ in $\Sigma_2$ originating from the long tail $R(s)\sim -1/N$ of 
the two-level correlation function. This is circumvented by fitting $\Sigma_2$ data with a polynomial 
in $\left< n\right>$ of degree $m$:
\begin{equation}
\label{chiap} 
\Sigma_{2}\left(\left< n\right>\right) \approx \chi_{0} + \chi \left<n\right> \ 
+ \sum_{k=2}^{m}\chi_{k}\left<n\right>^{k}
\end{equation}
Then we identify the coefficient of the {\em linear} term as $\chi$ and $\chi_0$ is 
the {\em constant} term. Both these terms are expected to have only weak $N$-dependence. This is in
contrast to that of the coefficients $\chi_{k}\propto 1/N^{k-1}$ with $k>1$ that have a strong 
$N$-dependence which describe finite-size effects in $\Sigma_{2}$. By making the polynomial fit 
Eqn.(\ref{chiap}) we separate the finite-size effects in $\Sigma_{2}$ and the weak $N$-dependence 
in $\chi$ and in $\chi_0$. This weak $N$-dependence in $\chi$ and in $\chi_0$ could be of fundamental 
interest and needs further investigation.

\begin{figure}
\begin{center}
\includegraphics[scale=0.6]{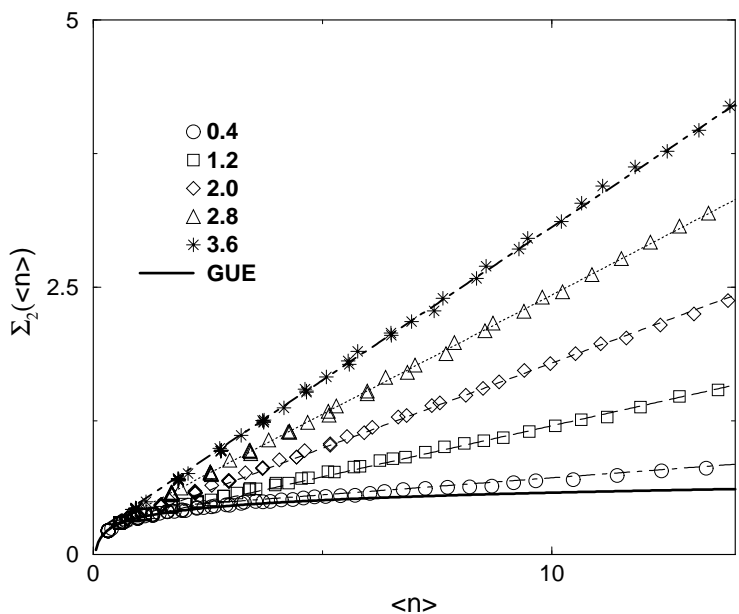}
\end{center}
\caption{\label{sigma2} The typical behavior of the level-number variance 
$\Sigma_{2}\left(\left< n\right> \right)$ for $N=400$ and for 
$b^{-1} = 0.4, 1.2, 2.0, 2.8$ and $3.6$. The lines through the data points 
correspond to $\chi_{0}\left( b \right)+\left(\frac{1}{4\pi b}\right)\left< n\right>$.
The value of $\chi_0$ is on average $N$ independent (see Fig.\ref{const0}). The linear 
behavior of $\Sigma_2$ is in perfect agreement with the theoretical prediction 
of \cite{AltZKS88,ChaKL96} (see Eqn.(\ref{pred})). For clarity in the numerical data
we skip every 3{\em rd} symbol.}
\end{figure}

\begin{figure}
\begin{center}
\includegraphics[scale=0.6]{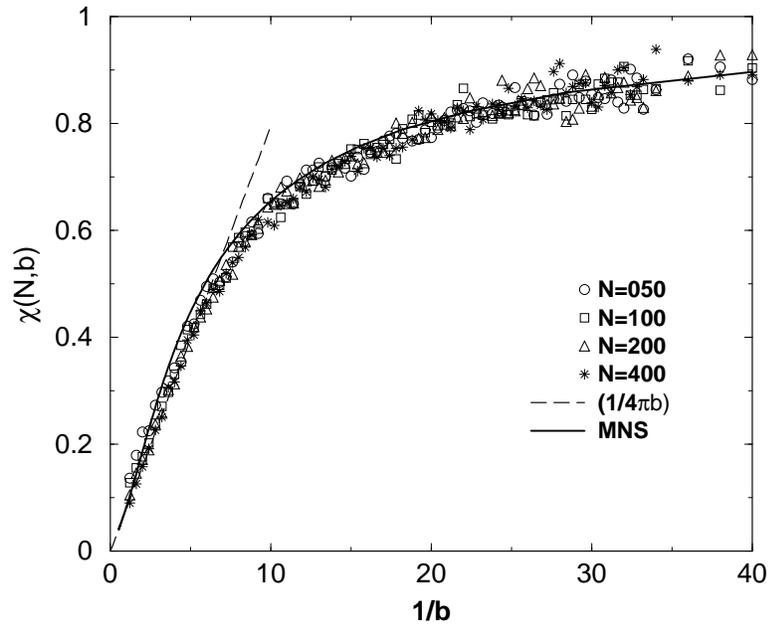}
\end{center}
\caption{\label{chi3}The spectral level compressibility $\chi$ as a function of $b^{-1}$ 
and $N$. The values of $\chi$ were calculated by a fit-polynomial of degree $m=3$. The 
solid line is the analytic formula (MNS) due to \cite{MosNS94} and the dashed line is 
Eqn.(\ref{pred}). $\chi$ is on average $N$ independent.}
\end{figure}

In Fig.\ref{sigma2} we plot $\Sigma_2$ for $N=400$ and $\Sigma_{2}\left(\left<n \right>\right)$ is 
linear as a function of $\left<n \right>$ as is expected for a critical ensemble. The lines through the 
data points are in agreement with the Poisson-like behavior {\em i.e.,} 
$\Sigma_2 \approx \chi \left< n \right> + \chi_{0}$. For large $b \gg 1/2\pi$ the value of 
$\chi$ is in good agreement with Eqn.(\ref{fract}) and Eqn.(\ref{pred}) that follows from 
Eqn.(\ref{fract}) and the correlation dimension $D_{2}$ found in Refs.\cite{Mir00,Kra96} : 
\begin{equation}
\label{pred}
\chi=1/4\pi b
\end{equation}
However, our results are in disagreement with Eqn.(\ref{fract}) for small $b \ll 1/2\pi$ 
\cite{MirE00,EveM00}. The analytic formula for $\chi$ (MNS in Fig.\ref{chi3}) that follows 
from \cite{MosNS94} for {\em all} values of $b$ reads:
\begin{equation}
\label{MNS}
\chi=\frac{d\ln\left({\cal I}\right)}{d\ln\left(z\right)}=\frac{\rm{PolyLog}\ 
\left[ -1/2,1-\exp \left(2\pi b\right)\right]}{\rm{PolyLog} \left[ +1/2,1-\exp \left(2\pi b\right)\right]}
\end{equation}
where $${\cal I}=\int_{0}^{\infty}\frac{dp}{\left(1+z^{-1}\exp\left(p^2\right)\right)}$$ and 
$\rm{PolyLog}\left[n,z\right]$ is an analytic continuation of $\sum_{k=1}^{\infty} z^{k}/k^{n}$.
This function can easily be computed by use of {\em mathematica}.

The main result of this work is that the values of $\chi$ (for modestly large $N<400$) of the 
critical ensemble due to \cite{MirFDQ96} agree unexpectedly well with the theoretical formula due
to \cite{MosNS94},
Eqn.(\ref{MNS}). This  agreement is, however, not perfect since the data points lie below
Eqn.(\ref{MNS}) in the range $3\le b^{-1}\le 10$, where the function $\chi\left(b^{-1}\right)$ in
Eqn.(\ref{MNS}) has a peculiar curved shape.

%
\begin{figure}
\begin{center}
\includegraphics[scale=0.6]{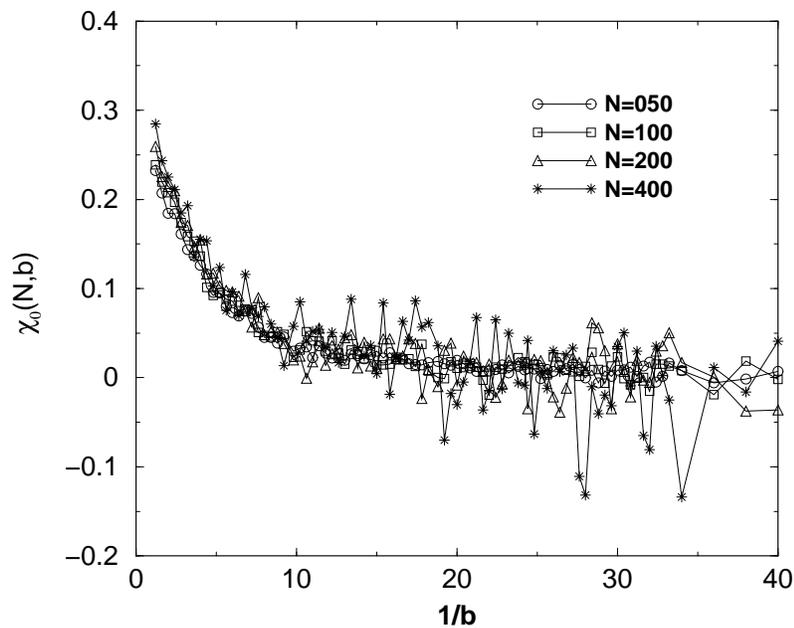}
\end{center}
\caption{\label{const0} The constant term $\chi_0(N,b)$ of the fit-polynomial with degree
$m=3$ in Eqn.(\ref{chiap}). $\chi_0$ decays to zero for values of $b^{-1}\le 10$. The data 
here is the same as in Fig.\ref{chi3}.}
\end{figure}
%
%
\section{Conclusion}
\label{conc}
We have been able to compute $\chi$ as a function of $b^{-1}$ for the GUE critical
ensemble. Our results are in agreement with the theoretical formula Eqn.(\ref{MNS}) 
except for the range $3 \le b^{-1} \le 10$ where the numerical data are below 
Eqn.(\ref{MNS}).

Furthermore, $\chi_0$ and $\chi$ are almost $N$-independent which is a signature of 
criticality. It is important to note that our simulations exclude the strong $N$-dependence 
of $\chi$ but they do not exclude the weak $\ln(N)$-dependence of $\chi$ which is not 
trivial to realize numerically. Recently, it has been shown analytically \cite{KYev} that 
in the CrRME the corrections to $\chi$ of order $b^{2}$ have a logarithmic $N$-term {\em i.e} 
$-b^{2}\ln(N)$ in the {\em crossover} between the GOE and the GUE. In both the 
GUE and the GOE this logarithmic factor is absent up to order $b^{2}$. It's presence is not 
excluded in higher $b$ corrections, {\em e.g} as $-b^{3}\ln(N)$. Such a logarithmic 
dependence is of fundamental interest. This is because the presence of $\ln(N)$-corrections 
would signal the {\it incomplete} criticality of the model. The logarithmic terms may drive the 
system to a different, stable, set of critical points where the $\chi$ may be drastically 
different from the MNS result. Unfortunately, the accuracy of our simulations is not enough to 
detect such a weak logarithmic dependence. This point needs further investigation. 

We note also that the constant term $\chi_0$ carries important information about CLS since it 
has a finite limit as $N \rightarrow \infty$. As is seen in figure \ref{const0}, the value of 
$\chi_0$ as a function of $b^{-1}$ decays to zero with increasing $b^{-1}$ which is (so far 
unnoticed) an important characteristic feature of CLS. 

In summary we have been able to compare via numerical simulations of the GUE critical
ensemble \cite{MirFDQ96} and the theoretical formula Eqn.(\ref{MNS}). The two results agree 
but in the region $3 \le b^{-1} \le 10$, the simulations are below the theoretical 
calculations. However, we emphasize that our results are preliminary and further work 
is in progress.
%
\section{Acknowledgments}
MLN would like to thank R. A. R\"{o}mer for encouraging discussions, funding from the 
Deutschen Forschungsgemeinschaft through the Sonderforschungsbereich 393 and the ICTP 
where this work was started. 
%
\section*{References}
%

\begin{thebibliography}{10}

\bibitem{Dys62}
F.~J. Dyson, J. Math. Phys. {\bf 3},  140  (1962).

\bibitem{Wig51}
E.~P. Wigner, Proc. Camb. Phil. Soc. {\bf 47},  790  (1951).

\bibitem{Meh90}
M.~L. Mehta, {\em Random Matrices} (Academic Press, Boston, 1990).

\bibitem{Haa92}
F. Haake, {\em Quantum Signatures of Chaos}, 2nd ed. (Springer, Berlin, 1992).

\bibitem{And58}
P.~W. Anderson, Phys. Rev. {\bf 109},  1492  (1958).

\bibitem{KraM97}
V.~E. Kravtsov and K.~A. Muttalib, Phys. Rev. Lett. {\bf 79},  1913  (1997).

\bibitem{Jan98}
M. Janssen, Phys. Rep. {\bf 295},  1  (1998).

\bibitem{Shkl} B.~I. Shklovskii, B. Shapiro, B.~R. Sears, P. Lambrianidis, and H.~B. Shore, Phys.Rev.B
{\bf 47}, 11487 (1993).

\bibitem{KLAA} V.E.Kravtsov, I.V.Lerner, B.L.Altshuler and A.G.Aronov, Phys.Rev.Lett.,{\bf 72},888
(1994).

\bibitem{MosNS94}
M. Moshe, H. Neuberger, and B. Shapiro, Phys. Rev. Lett. {\bf 73},  1497
  (1994).

\bibitem{MirFDQ96}
A.~D. Mirlin {\it et~al.}, Phys. Rev. E {\bf 54},  3221  (1996).

\bibitem{MutCIN93}
K.~A. Muttalib, Y. Chen, M.~E.~H. Ismail, and V.~N. Nicopoulos, Phys. Rev.
  Lett. {\bf 71},  471  (1993).

\bibitem{Mir00}
A.~D. Mirlin, Phys. Rep. {\bf 326},  259  (2000).

\bibitem{BogBP97}
E. Bogomolny, O. Bohigas, J. Keating, and M.~P. Plato, Phys. Rev. E {\bf 55},
  6707  (1997).

\bibitem{GarV00}
A.~M. Garcia-Garcia and J.~J.~M. Verbaarschot, Nucl. Phys. B {\bf 586},  668
  (2000).

\bibitem{BleCM94}
C. Blecken, Y. Chen, and K.~A. Muttalib, J. Phys. A: Math. Gen. {\bf 27},  L563
   (1994).

\bibitem{GarV02}
A.~M. Garcia-Garcia and J.~J.~M. Verbaarschot,   (2002), {ArXiv}:
  cond-mat/0204151.

\bibitem{ChaKL96}
J. Chalker, V. Kravtsov, and I. Lerner, Pis'ma Zh. Eksp. Teor. Fiz. {\bf 64},
  355  (1996), [Sov. Phys. JETP Lett.\ {\bf 64}, 386 (1996)].

\bibitem{KTs} V.E.Kravtsov and A.M.Tsvelik. Phys. Rev. B., 
{\bf 62}, 9888 (2000).

\bibitem{Nish} S.M.Nishigaki, Phys.Rev.E {\bf 59}, 2853 (1999).

\bibitem{MirE00}
A.~D. Mirlin and F. Evers, Phys. Rev. B {\bf 62},  7920  (2000).

\bibitem{EveM00}
F. Evers and A.~D. Mirlin, Phys. Rev. Lett. {\bf 84},  3690  (2000).

\bibitem{Can96}
C.~M. Canali, Phys. Rev. B {\bf 53},  3713  (1996).

\bibitem{AltZKS88}
B.~L. Altshuler, I.~K. Zharekeshev, A.~A. Kotochigova, and B.~I. Shklovskii,
  Pis'ma Zh. Eksp. Teor. Fiz. {\bf 94},  343  (1988), [Sov. Phys. JETP Lett.\
  {\bf 67}, 625 (1988)].

\bibitem{Kra96}
V.~E. Kravtsov, Proceedings of the Correlated Fermions and Transport in
  Mesoscopic Systems, Moriond Conference, {Les, Arcs}  (1996), {ArXiv}:
  cond-mat/9603166.

\bibitem{ChaLS96}
J.~T. Chalker, I.~V. Lerner, and R.~A. Smith, Phys. Rev. Lett. {\bf 77},  554
  (1996).

\bibitem{KYev} O.~M. Yevtushenko and V.~E. Kravtsov (unpublished)

\end{thebibliography}

\end{document}